# Thermal building simulation and computer generation of nodal models

H. BOYER, J.P. CHABRIAT, B. GRONDIN-PEREZ, C. TOURRAND

*Université de la Réunion, Faculté des Sciences, Laboratoire de Génie Industriel, BP 7151,*

*15 avenue René Cassin, 97705 Saint-Denis Cedex, FRANCE (DOM)*

J. BRAU

*INSA de Lyon, CETHIL / Equipement de l'Habitat, Bât. 307, 20 rue Albert Einstein,*

*69621 Villeurbanne Cedex, FRANCE*

*The designer's preoccupation to reduce the energy needs and get a better thermal quality of ambiances helped in the development of several packages simulating the dynamic behaviour of buildings. This paper shows the adaptation of a method of thermal analysis, the nodal analysis, linked to the case of building's thermal behaviour. We take successively an interest in the case of conduction into a wall, in the coupling with superficial exchanges and finally in the constitution of thermal state models of the building. Big variations existing from one building to another, it's necessary to build the thermal model from the building description. This article shows the chosen method in the case of our thermal simulation program for buildings, CODYRUN.*

# I INTRODUCTION

The nodal analysis is a powerful method of investigation in the thermal analysis of systems. It has been used in several branches such as solar energy systems [1], micro-electronics [2] or also the spatial field [3]. We will gradually use this approach in the domain of building's physics and we'll interest ourselves in the automatic generation of nodal models.

Considering the thermal behaviour of a building, its thermal state is determined by the continuous field of temperatures, concerning all points included in the physical limits of the building. The constitution of a reduced model (shown in fig. 1), with a finite number of temperatures, is possible by assuming some simplifications. So, we consider mono-dimensional heat conduction in walls and well mixed air volumes of each thermal zone.

*Fig. 1 : Thermal discretisation of a building*

Having explained the choosen level of modelisation, our objective was to build a dedicated software tool usable by researchers and suitable for professionals. This last aspect took us not to use general purpose simulation softwares, as TRNSYS [4].

More precisely, *CODYRUN* is a multizone software integrating both natural ventilation and moisture transfers, developed on a PC micro computer. One of its most interesting aspects is to offer the expert thermician a wide range of choices between different heat transfer models and meteorological reconstitution parameters models. The aim may be to realize studies of the software sensitivity to these different models (in regard to precision or calculation time), in order to choose the ones that should be integrated in a suitable conception tool for a particular climate. Because of the amount of information necessary to describe one building, we've built a user friendly front-end, based on Microsoft Windows environment We've had to develop a specific program because analysis of existing software didn't allow us to respect our needs. Meanwhile, in order to compare *CODYRUN* to other programs, let's say that the zones coupling approach (detailled in this paper) is similar to the one used in ESP [5] and that the pressure airflow model, integrating large openings, is quite the same as TARP [6].

## II  NODAL ANALYSIS APPLIED TO HEAT CONDUCTION

For an envelope's wall, we suppose that we have two temperatures as conditions to surface's limits. In the hypothesis of monodimensional conductive transfers, the study's frame is then cut into a determinated number of elements supposed in each moment of a uniform temperature (3 in the figure below).

*Fig. 2 : Wall spatial discretisation*

The formal analogy between the equation of conduction made by Fourier and the Ohm's law in an electrical conductor, that is to say

$$\vec{\varphi} = -\lambda \ \mathbf{grad} \ T, \ \vec{j} = \frac{1}{\rho} \mathbf{grad} \ V \qquad (1a,b)$$

induces the correspondence between the following groupings :

$$(\varphi, j) \ , \ (\lambda, \rho^{-1}) \ and \ (T, V)$$

Then, the transposition of the thermal problem of conduction into an electrical problem is known under the name of thermo-electrical analogy. Working this analogy, the nodal method leads to the setting up of an electrical network as the following.

*Figure 3 : Associated electrical network*

The nodes, which in an electrical meaning symbolize equipotentials, correspond then to isotherm lines. Applied to the monodimensional conduction in a wall, those isothermal elements are isothermal slices sprung from our discretisation hypothesis. Then, those nodes are linked to each other by the analogical resistance of the physical layer of the wall which divides them. Therefore, each of those nodes are getting an electrical capacitor, traducing the thermal storage of the corresponding wall's part, and allowing in this the traduction of the thermal inertia effects. Then, by applying the conservation law of currents in each point (Kirchoff's law), the solution of the problem or the partial derivatives amounts to the one of an algebra-differential system. If the number of dicretisation node is *n*, as taken into account capacities and thermal conductances, the obtained system is the following,

$$\begin{cases} C_1 \dfrac{dT_1}{dt} = K_{0,1}(T_{si} - T_1) + K_{1,2}(T_2 - T_1) \\ C_2 \dfrac{dT_2}{dt} = K_{1,2}(T_1 - T_2) + K_{2,3}(T_3 - T_2) \\ .... \\ C_n \dfrac{dT_n}{dt} = K_{n-1,n}(T_{n-1} - T_n) + K_{n,n+1}(T_{se} - T_n) \end{cases} \quad (2)$$

,considering that the number *0* is attached to the inside surface node and *n+1* to the outside surface one. The former system can be also written under a matricial form

$$\begin{bmatrix} C_1 & 0 & 0 & 0 \\ 0 & C_2 & 0 & 0 \\ 0 & 0 & ... & 0 \\ 0 & 0 & 0 & C_n \end{bmatrix} \begin{Bmatrix} \dot{T}_1 \\ \dot{T}_2 \\ ... \\ \dot{T}_n \end{Bmatrix} = \begin{bmatrix} -K_{0,1} & 0 & 0 & 0 \\ K_{1,2} & -K_{1,2} & 0 & 0 \\ ... & ... & ... & ... \\ 0 & 0 & 0 & -K_{n,n+1} \end{bmatrix} \begin{Bmatrix} T_1 \\ T_2 \\ ... \\ T_n \end{Bmatrix} + \quad (3)$$

$$\begin{bmatrix} K_{0,1} & 0 & 0 & 0 \\ 0 & 0 & 0 & 0 \\ 0 & 0 & 0 & 0 \\ 0 & K_{n,n+1} & 0 & 0 \end{bmatrix} \begin{Bmatrix} T_{si} \\ T_{se} \\ 0 \\ 0 \end{Bmatrix}$$

This matricial differential equation forms the state equation of evolution of the system, taking into account its discretisation. Using the matricial formalism, the former system is written under the form

$$C_w \dot{T}_w = A_w T_w + B_w \quad (4)$$

the index *w* used for walls and windows. The same method can be found in two-dimensional transient heat conduction [7].

### III COUPLING WITH SUPERFICIAL EXCHANGES

For a wall separating the interior of the building from the exterior, the following figure makes explicit the boundary conditions.

*Figure 4 : Boundary conditions*

$\varphi_{swe}$ and $\varphi_{swi}$ are flux densities of short wave length, respectively to the exterior and the interior. Those flux densities are averaged for a given time step, and depend on meteorological conditions. $\varphi_{ce}$ and $\varphi_{ci}$ concern the convection's terms, respectively exterior and interior. In the hypothesis of Newtonian models, by introducing linear exchange coefficients $h_{ce}$ and $h_{ci}$, those flux densities are written :

$$\begin{aligned}\varphi_{ce} &= h_{ce}\ (T_{se} - T_{ae}) \\ \varphi_{ci} &= h_{ci}\ (T_{si} - T_{ai})\end{aligned} \quad (5a,b)$$

$T_{ae}$ and $T_{ai}$ being respectively dry-bulb temperature of outside and inside air.

$\varphi_{lwe}$ and $\varphi_{lwi}$ correspond to the radiative exchanges of long wave length. Classicaly linearized in the range of building temperatures, we write then :

$$\begin{aligned}\varphi_{lwe} &= h_{re}\ (T_{sky} - T_{se}) \\ \varphi_{lwi} &= h_{ri}\ (T_{si} - T_{rm})\end{aligned} \quad (6a,b)$$

To simplify the formalism of radiative tranfers, we introduced outside the sky radiant temperature ($T_{sky}$) and inside mean radiant temperature ($T_{rm}$).

**IV SINGLE ZONE BUILDING MODEL**

In a given building, this one is made up with a certain number of rooms, walls or also glass windows. The following figure makes explicit a building's configuration, as obtained by assembling elementary components, such as walls, doors, glass windows ... :

*Fig. 5 : Component assembly of a building*

Then, the physical model of the building is obtained by assembling thermal models of each element such as walls and glass windows. To close the problem, we add the thermo-convective balance equation of dry-bulb air node and the radiative balance equation of the inside mean radiant temperature node. To simplify our discussion, we'll suppose that the heat conduction is treated with the help of a model constituted of a thermal resistance and 2 capacitors, said "R2C" model [8]. Therefore, there is no internal node inside the wall. The mathematical traduction of the thermal model of the building is consequently a linear system, including a certain number of equations of type *7* and *8*, and one equation of type *9* and another one of type *10*.

$$C_{si} \frac{dT_{si}}{dt} = h_{ci}(T_{ai} - T_{si}) + h_{ri}(T_{rm} - T_{si}) + K(T_{se} - T_{si}) + \varphi_{swi} \quad (7)$$

$$C_{se} \frac{dT_{se}}{dt} = h_{ce}(T_{ae} - T_{se}) + h_{re}(T_{sky} - T_{se}) + K(T_{si} - T_{se}) + \varphi_{swe} \quad (8)$$

$$C_{ai} \frac{dT_{ai}}{dt} = \sum_{j=1}^{Nw} h_{ci}(T_{ai} - T_{si(j)}) + c\,\dot{Q}(T_{ae} - T_{ai}) \quad (9)$$

$$0 = \sum_{j=1}^{Nw} h_{ri}\,A_j\,(T_{si(j)} - T_{rm}) \quad (10)$$

Equations *(7)* and *(8)* traduce the respective thermal balance of the nodes of inside and outside surfaces. $N_W$ designating the number of walls of the envelope, equation *(9)* is the one of the thermo-convective balance of the dry-bulb inside air temperature, taking into account airflow between inside and outside. *(10)* is the equation of the radiative balance of mean radiant temperature node.

This system can be written again under the form $\mathbf{C}_z\,\dot{\mathbf{T}}_z = \mathbf{A}_z\,\mathbf{T}_z + \mathbf{B}_z$, by adding to the state vector of the wall temperatures the indoor dry-bulb temperature ($T_{ai}$) and the mean radiant temperature ($T_{rm}$).

$$\begin{bmatrix} \mathbf{C}_w & \\ & C_{ai} \\ & & 0 \end{bmatrix} \begin{bmatrix} \dot{\mathbf{T}}_w \\ \dot{T}_{ai} \\ \dot{T}_{rm} \end{bmatrix} = \begin{bmatrix} \mathbf{A}_w & \begin{matrix} h_{ci} & h_{ri} \\ h_{ci} & h_{ri} \end{matrix} \\ \hline \cdots\cdots\cdots\cdots \end{bmatrix} \begin{bmatrix} \mathbf{T}_w \\ T_{ai} \\ T_{rm} \end{bmatrix} + \begin{bmatrix} \mathbf{B}_w \\ \hline \cdots \end{bmatrix} \quad (11)$$

We coloured in grey the elements of the state system only corresponding to the phenomenon of conduction. In the matrix $\mathbf{A}_z$, the terms on the right of $\mathbf{A}_w$ traduce the coupling of the three transfer modes on an envelope surfacic node, located indoor. Single zone softwares, as *CODYBA* [9], are based on solution of this linear system.

To establish automatically this state equation, we propose a splitting-up of the $\mathbf{A}_z$ matrix into 12 elementary matrixes. Each of those elementary objects concerns particular terms. The same, the filling up to each time step of the B vector is facilitated by its decomposition into 15 elementary vectors. Thus,

$$[C]\left\{\frac{dT}{dt}\right\} =$$
$$[A_{cond} + A_{cvi} + A_{cve} + A_{lwe} + ... + {}_{connex}]\{T\} + \quad (12)$$
$$\{B_{swi} + B_{swe} + B_{lwi} + ... + B_{connex}\}$$

$A_{cond}$ includes the terms linked to heat conduction in the walls and $A_{cvi}$, the terms linked to the interior linearised convective exchanges. $B_{swe}$ is the vector of exterior radiative flux densities. $A_{connex}$ and $B_{connex}$ will be explained further.

Then, a finite differences scheme of the time variable is used for numerical resolution. The determination of the temperatures field is made in two steps, as follows :

*Fig. 6 : Detail of one zone solution*

In case of air conditionning, using the same formalism, the problem of finding an optimal control scheme can be explored [10].

**V  MULTIZONE CASE**

Considerating a multizone building, the different zones' temperatures (principal variables) are linked together through heat conduction and the air mouvement. By adopting the same proceeding as before, for the whole building, the method leads to a linear system of big dimension. With an adequate numbering, it's possible to make the matricial state equation of each zone appeared, and also the terms corresponding to the couplings. For a building requiring two zones indexed respectively by *1* and *2*, the state equation appears then this way :

$$\begin{bmatrix} C_1 & \\ & C_2 \end{bmatrix} \begin{bmatrix} \dot{T}_1 \\ \dot{T}_2 \end{bmatrix} = \begin{bmatrix} A_1 & \times \times \\ \times \times & A_2 \end{bmatrix} \begin{bmatrix} T_1 \\ T_2 \end{bmatrix} + \begin{bmatrix} B_1 \\ B_2 \end{bmatrix} \quad (13)$$

And more than the numerical incidences of the resolution of such a linear system (the matrix **A** is quite empty), this approach doesn't put up very well with our objective of multimodel tool [11]. We preferred to use an approach which we call "connected", insuring the physical coupling of the zones between themselves by a process of iterative resolution. So, when calculating the temperatures associated to a zone, the characteristic temperatures (dry-bulb and mean radiant) of the other zones are supposed known. The following flow chart traduces the thermal coupling of the zones :

*Fig. 7 : Coupling flowchart*

Other methods can be found, as [12], based on topological simplification of electrical networks. The particularity of the chosen coupling method [13] is the presence of a common part of each elementary model. Then, the common part, the "recovery place", allows to couple models to each other in a simple manner. From the view point of our calculation's organisation, the coupling of the considered zone with the other zones is insured thanks to the filling of particular matrix and vector, **A**$_{connex}$ and **B**$_{connex}$.

*Fig 8 : Building model assembly*

The necessary number of iterations to reach convergence depends of course on the intensity of thermal zones couplings. Let's consider the case of two rooms with very different thermal behaviours, for example one warmed and the other not. The number of iterations that we can presume important (and penalizing for our coupling's mode) will be the reduced weak because the isolation between the two rooms would have been correctly made, which is legitimate to rely on.

This paragraph makes explicit the method in the case of a simplified two-zone building, supposed without ceiling and floor and whose view plan view is given in figure 9. The chosen conductive model in each wall is of type *R2C*. The discretisation leads us then to arrange 18 temperature nodes in this building.

*Fig. 9 : Simple two-zone building.*

In the case of direct resolution of the coupled system, the state vector is made of nodes temperatures from $T_1$ to $T_{18}$. Then, the system to be solved is of dimension 18. Considering the definition of a zone model, the state vector of the first zone (on the left ) is composed with ($T_1$, $T_2$, $T_3$, $T_4$, $T_5$, $T_6$, $T_7$, $T_8$, $T_{15}$, $T_{16}$), while the one of the second zone concerns temperatures ($T_7$, $T_8$, $T_9$, $T_{10}$, $T_{11}$, $T_{12}$, $T_{13}$, $T_{14}$, $T_{17}$, $T_{18}$). We notice that $T_7$ and $T_8$ belong to each state vectors, because the two corresponding discretisation nodes belong to the recovery place (hatched on the diagram). Each zone sub-system is of dimension 10.

Then, for the first zone, an illustration of a part of the decomposition of the former paragraph can be given. To simplify the writing, the internal convective transfers are integrated with a linear model with a constant exchange coefficient $h_{ci}$, as for the internal radiative exchanges (with $h_{ri}$).

$$A_{cond} = \begin{bmatrix} -K_{1,2} & K_{1,2} & 0 & 0 & 0 & 0 & 0 & 0 & 0 & 0 \\ K_{1,2} & -K_{1,2} & 0 & 0 & 0 & 0 & 0 & 0 & 0 & 0 \\ 0 & 0 & -K_{3,4} & K_{3,4} & 0 & 0 & 0 & 0 & 0 & 0 \\ 0 & 0 & K_{3,4} & -K_{3,4} & 0 & 0 & 0 & 0 & 0 & 0 \\ 0 & 0 & 0 & 0 & -K_{5,6} & K_{5,6} & 0 & 0 & 0 & 0 \\ 0 & 0 & 0 & 0 & K_{5,6} & -K_{5,6} & 0 & 0 & 0 & 0 \\ 0 & 0 & 0 & 0 & 0 & 0 & -K_{7,8} & K_{7,8} & 0 & 0 \\ 0 & 0 & 0 & 0 & 0 & 0 & K_{7,8} & -K_{7,8} & 0 & 0 \\ 0 & 0 & 0 & 0 & 0 & 0 & 0 & 0 & 0 & 0 \\ 0 & 0 & 0 & 0 & 0 & 0 & 0 & 0 & 0 & 0 \end{bmatrix}$$

$$A_{cvi} = \begin{bmatrix} 0 & 0 & 0 & 0 & 0 & 0 & 0 & 0 & 0 & 0 \\ 0 & -h_{ci} & 0 & 0 & 0 & 0 & 0 & 0 & h_{ci} & 0 \\ 0 & 0 & 0 & 0 & 0 & 0 & 0 & 0 & 0 & 0 \\ 0 & 0 & 0 & -h_{ci} & 0 & 0 & 0 & 0 & h_{ci} & 0 \\ 0 & 0 & 0 & 0 & -h_{ci} & 0 & 0 & 0 & h_{ci} & 0 \\ 0 & 0 & 0 & 0 & 0 & 0 & 0 & 0 & 0 & 0 \\ 0 & 0 & 0 & 0 & 0 & 0 & -h_{ci} & 0 & h_{ci} & 0 \\ 0 & 0 & 0 & 0 & 0 & 0 & 0 & 0 & 0 & 0 \\ 0 & 0 & 0 & 0 & 0 & 0 & 0 & 0 & 0 & 0 \\ 0 & 0 & 0 & 0 & 0 & 0 & 0 & 0 & 0 & 0 \end{bmatrix}$$

$$A_{connex} = \begin{bmatrix} 0 & 0 & 0 & 0 & 0 & 0 & 0 & 0 & 0 & 0 \\ 0 & 0 & 0 & 0 & 0 & 0 & 0 & 0 & 0 & 0 \\ 0 & 0 & 0 & 0 & 0 & 0 & 0 & 0 & 0 & 0 \\ 0 & 0 & 0 & 0 & 0 & 0 & 0 & 0 & 0 & 0 \\ 0 & 0 & 0 & 0 & 0 & 0 & 0 & 0 & 0 & 0 \\ 0 & 0 & 0 & 0 & 0 & 0 & 0 & 0 & 0 & 0 \\ 0 & 0 & 0 & 0 & 0 & 0 & 0 & 0 & 0 & 0 \\ 0 & 0 & 0 & 0 & 0 & 0 & 0 & -(h_{ci}+h_{ri}) & 0 & 0 \\ 0 & 0 & 0 & 0 & 0 & 0 & 0 & 0 & 0 & 0 \\ 0 & 0 & 0 & 0 & 0 & 0 & 0 & 0 & 0 & 0 \end{bmatrix}$$

$$B_{connex} = \begin{Bmatrix} 0 \\ 0 \\ 0 \\ 0 \\ 0 \\ 0 \\ 0 \\ h_{ci}T_{17} + h_{ri}T_{18} \\ 0 \\ 0 \end{Bmatrix}$$

**VI  DYNAMIC BUILDING NODAL ANALYSIS**

A given building is composed with a certain number of rooms, walls or also glass-windows. At the beginning of this variable description, the building's decomposition into a certain number of zones is a simulation parameter. Taking into account our scheme of iterative connexion, our principal problem is the establishment of the state equation of each zone.

*Fig 10 : From building description to its model*

**1 ) The nodal structure :**

To fill up each of those mathematical objects (i.e. C, A and B) from the description files of the building (walls, glass-windows, ...), we build a data structure constituted of information fields attached to each discretisation node. Firstly, an incremental number is attached to each node, to index our structure. The building description bringing in a certain number of walls and glass-windows, knowing the number of nodes constituting the discretisation of each entity, all the walls and glass-windows nodes can be numbered. Therefore, the building splitting up into thermal zones induces the setting of two nodes of temperature by zone (dry-bulb air and mean radiant). The following figure explains our proceeding.

Fig. 11 : *The nodal structure generation*

A certain number of information fields are connected to a node, traducing for instance the allocation of a node to a zone or also the topology of the global electrical network associated to the building. Considering the objective to be reached, we have been induced to impute a type to each node. Indeed, relatively to the equations, the nodes are concerned by different phenomenons. For instance, a wall node is going to concern terms of heat conduction. This same node, depending on its location, can also concern convective process. On the external face of the envelope's wall, the surfacic node is concerned by outdoor radiative and convective exchanges. Then, it appears necessary to attribute a type to each node. To record those types, we just had to consider each object taken from the decomposition of the state equation and wonder : *what is the characteristic of the nodes concerning the filling up of this object ?* Answering previous question, the table below gives in its first column the encountered types of nodes and in the second one a reference number that we'll find again on the next figure.

| Type | Ref. |
|---|---|
| *Outdoor surfacic node of outside wall* | *1* |
| *Indoor surfacic node of outside wall* | *2* |
| *Internal node of outside wall* | *3* |
| *Outdoor surfacic node of outside glass-window* | *4* |
| *Indoor surfacic node of outside glass-window* | *5* |
| *Surfacic node of internal wall* | *6* |
| *Internal node of internal wall* | *7* |
| *Surfacic node of vertical interzone wall* | *8* |
| *Internal node of interzone wall* | *9* |
| *Surfacic node of ground wall* | *10* |
| *Internal node of ground wall* | *11* |
| *Terminal node of ground wall* | *12* |
| *Surfacic below node of horizontal interzone wall* | *13* |
| *Surfacic under node of horizontal interzone wall* | *14* |
| *Surfacic node of interzone glass-window* | *15* |
| *Dry indoor air temperature* | *16* |
| *Radiant mean temperature* | *17* |

Table 1 : *Encountered types of nodes*

On the building seen in section in figure 12, we have only represented (with a white circle) a few nodes of thermal discretisation. For these nodes, we've indicated the number taken from the former table corresponding to its type.

*Fig. 12: Types of nodes.*

The multizone feature requires the nodes to be located with regard to the chosen zones split-up. Indeed, all the discretisation nodes of "recovery place" belong simultaneously to two zones. Then, the structure will include two zones' numbers. More than the incremental numbering of nodes, each node gets a number relating to a zone. For one of the zones to which the node belongs, this relative number is equal to the line number of its balance equation in this zone model (system which put into the matrix form composes the state equation).

Therefore, each node will own, as well as its absolute type, another type relating to the zone. Indeed, in our connexion process of the zones together, certain nodes play a particular part, so that in their balance equation interfere terms concerning the adjacent zone (connexion nodes). The figure 13, of a network 3R2C of an interzone wall, details the nodes' structure.

*Fig. 13 : Interzone partition and associated data structure*

By taking the example of the former figure, when writing the balance equations of the first zone nodes, the node *l* is concerned by the dry-bulb and mean radiant temperature of the second zone. Then, we say that this node is connexion node for the zone 1 (its relative type regard to zone 1 is marked in the structure with *1* and with a *0* regard to the zone 2). To link with the objects of the evolution equation of a zone, such a node (wall or interzone glass-window's surface) will concern the filling up of $\mathbf{A}_{connex}$ and $\mathbf{B}_{connex}$, provided that this node owns the relative type "connexion node" in regard to the considered zone.

The nodal discretisation imputes also a capacitor to each node, with a value that can be null (glass windows' nodes or surface's nodes of walls in case of conductive models different from the *R2C*). The calculation of these capacitors (and resistors) is performed in a step before simulation and recorded in the nodes' structure.

Therefore, the nodes of a same wall are linked to each other by conductances. In order to have all the necessary information to the reconstitution of the global electrical network, it's necessary to record for each node the associated conductance (*K*) and the incremental number of the node to which it's connected. The composition of this structure understood, a module of *CODYRUN* undertakes its generation, from the hierarchical description of the building, before starting the simulation. This module is performing the physical discretisation of the building and so, treats sequentially each zone, each wall and each glass window.
For example, if we consider again the two-zone building of figure 9, assuming thet all the conductances are equal to *K*, the nodes' structure associated to this building is the following one :

*Fig 14 : Nodes' structure corresponding to fig. 9 building*

We have to note that the size of this structure can quickly become significant, a building having most of the time several zones and for each zone several walls and glass-windows. This structure's size being linked to the dimensions of the linear systems to be solved, the notion of calculation time mustn't be overlooked [14].

## 2 ) From the node's structure to the mathematical model

For a given building, when the nodes' structure is established, it's easy to fill up each element of the mathematical model. Indeed, we have just to sweep the nodes structure and make the attribution of the concerned terms. As for instance to fill up the matrix $\mathbf{A}_{cond}$ of a zone. By sweeping of the structure, it's possible to choose (with the help of their type) only the wall and glass-window nodes (absolute types not equal to 16 and 17 on figure 12), belonging in addition to the considered zone, indexed by $nZ$. For a node, a necessary condition to check is then that almost one of the zone allocation number (indicated in the nodes' structure as *zones' numbers*) is equal to $nZ$. As $n$ the absolute number of theses nodes. Then, the node $n$ owns in the zone $nZ$ a relative number $i$ (indicated in the nodes'structure as *relative zone number)*.

The structure informs us also about the absolute number of the node $m$ to whom $n$ is connected, and about the value of the conductance $K_{n,m}$. This node $m$ owns also a relative number in the zone $nZ$, that is to say $j$. To fill up the matrix $\mathrm{A}_{cond}$ with the terms corresponding to the conductance between nodes $n$ and $m$ is performed by affecting $K_{n,m}$ to $\mathbf{A}_{cond}[i,j]$ and adding $-K_{n,m}$ to $\mathbf{A}_{cond}[i,j]$. With this method, using fig. 14, it's easy to find again $\mathbf{A}_{cond}$, $\mathbf{A}_{cvi}$, $\mathbf{A}_{connex}$ and $\mathbf{B}_{connex}$ given for the first zone of building in fig 9.

The same proceeding is used to fill up all the other objects of the evolution equation, by using meteorological data of the current time step for the vectors linked to the sollicitations ($\mathbf{B}_{swe}$, $\mathbf{B}_{cve}$, ...) and estimate values of adjacent temperatures for the connexion terms ($\mathbf{B}_{connex}$).

## VII  CONCLUSION

The physical principle of nodal analysis associated to a sequential program of generation of model showed its potential interest in the constitution of building thermal model. Another aspect of to this method concerns the multiple model approach that we had. From other considerations [11], we though it was interesting to be able in a given building to refine the discretisation of one or several walls. Then, its necessary to inform about the chosen conductive model during the description. By taking into account this conductive model during the writing of the building model, this information "model" is then associated (and recorded) with the description's information. Then, according again to figure 11, a more detailed corresponding node's structure can be easily generated.

The interest of the proceeding and intermediary stage of the generation of nodes' structure still appears clearly in consideration of future development of specific tools to *CODYRUN*. This structure, which size can be significant, can be the object of a processing (for its reduction) with skiful rules (automatic choice of zones depending on the objective, optimal choice of conductive models, ......). The easiest example is the one of the reduction of the building splitting-up into zones. A multizone building described, it can be interesting to be able to simulate the building considering a number of zones being lower to the one scheduled during the description. A reason can be for instance the quick disposal of certain elements of answer, as an estimation of energy needs on a long period. To avoid describing once more the building, writing a module sensibly changing the nodes structure is enough, before performing the desired simulation. Another application is the case of architectural sketch recognition [15], in which the intermediary stage of the nodal structure is a convenient stage between recognized entities (zones, walls, ...) and the thermal model generation.

*Acknowledgement* - The financial contribution of Reunion Island delegation of A.D.E.M.E. (*Agence De l'Environnement et de la Maîtrise de l'Energie*) to this study is gratefully acknowledged.

**NOMENCLATURE**

- **A** square matrix
- *A* area
- **B** vector
- **C** diagonal matrix
- *C* thermal capacity
- *A* area
- *c* specific heat
- *h* heat transfer coefficient
- *K* thermal conductivity
- $\varphi$ radiation flux density
- $\dot{Q}$ mass flow rate
- *j* electrical intensity density
- $\lambda$ thermal conductivity
- $\rho$ electrical resistivity or volumic mass
- *T* temperature
- *V* voltage

Subscripts :

- *ae*     air ouside
- *ai*     air inside
- *se*     surface outside
- *si*     surface inside
- *ce*     outside convection
- *ci*     inside convection
- *lwe*     long wave outside
- *re*     outside radiation

| | |
|---|---|
| *ri* | *inside radiation* |
| *rm* | *inside radiant mean* |
| *swe* | *short wave outside* |
| *swi* | *short wave inside* |
| *w* | *wall* |
| *z* | *zone* |

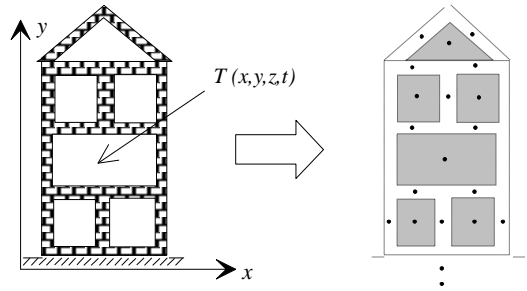

*Fig. 1 : Thermal discretisation of a building*

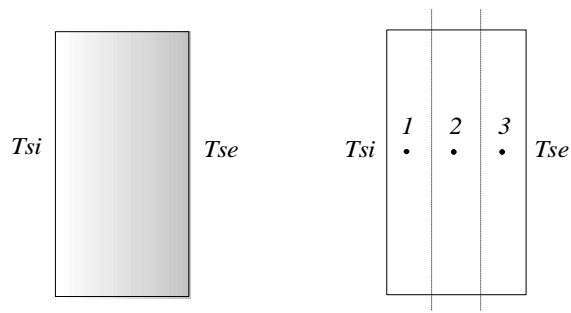

*Fig. 2 : Wall spatial discretisation*

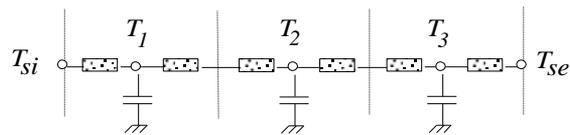

*Figure 3 : Associated electrical network*

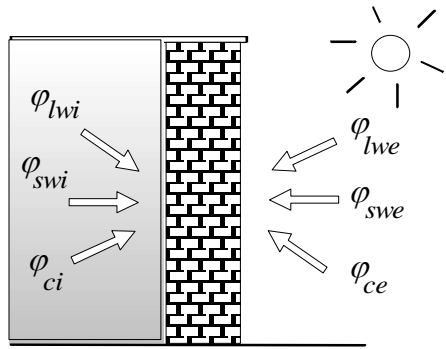

*Figure 4 : Boundary conditions*

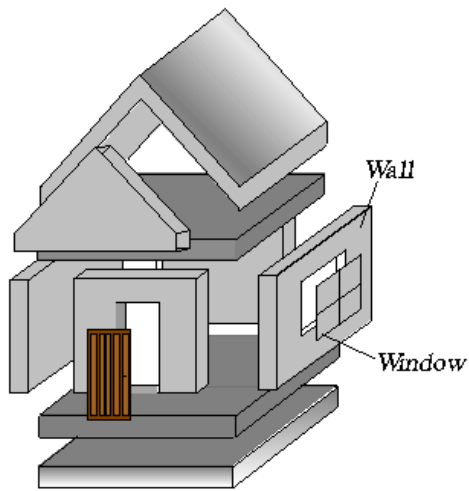

*Fig. 5 : Component assembly of a building*

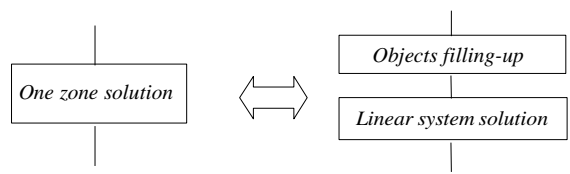

*Fig. 6 : Detail of one zone solution*

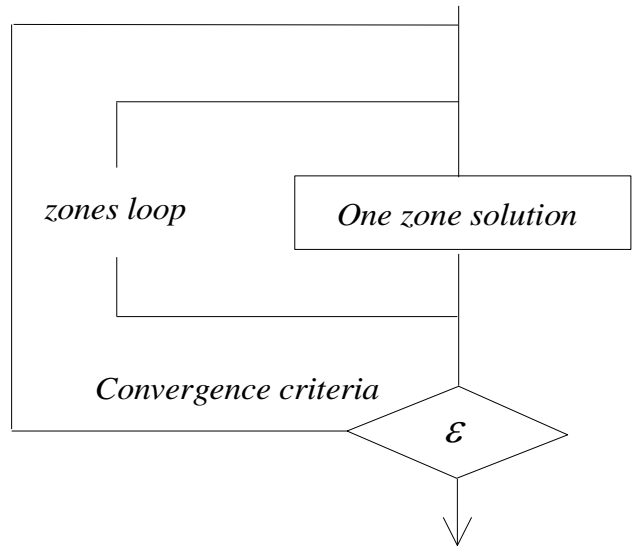

*Fig. 7 : Coupling flowchart*

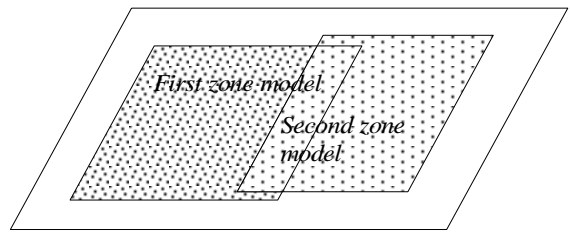

*Fig 8 : Building model assembly*

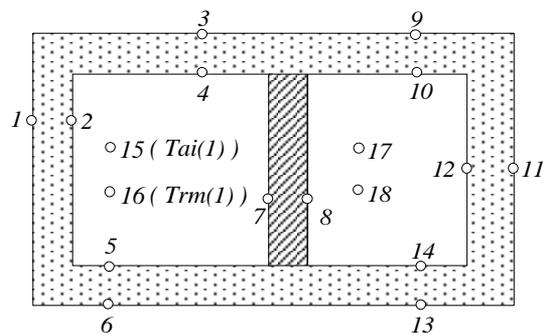

*Fig. 9 : Simple two-zone building*

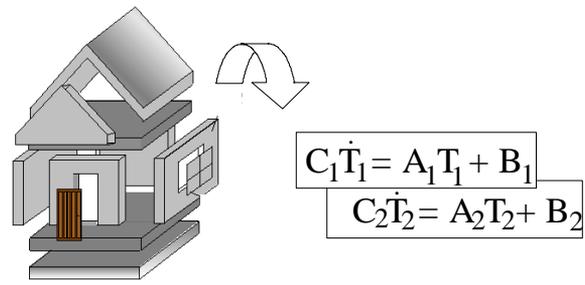

*Fig 10 : From building description to its model*

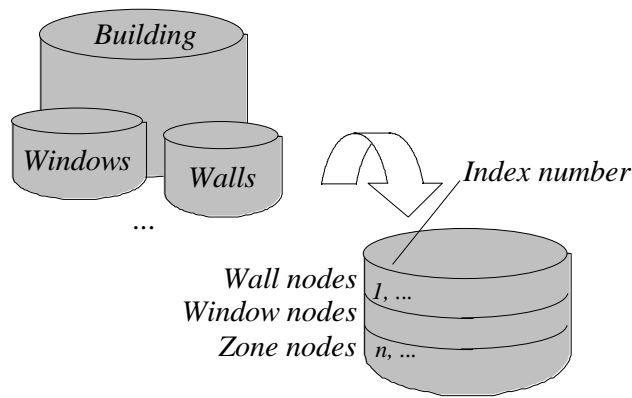

Fig. 11 : *The nodal structure generation*

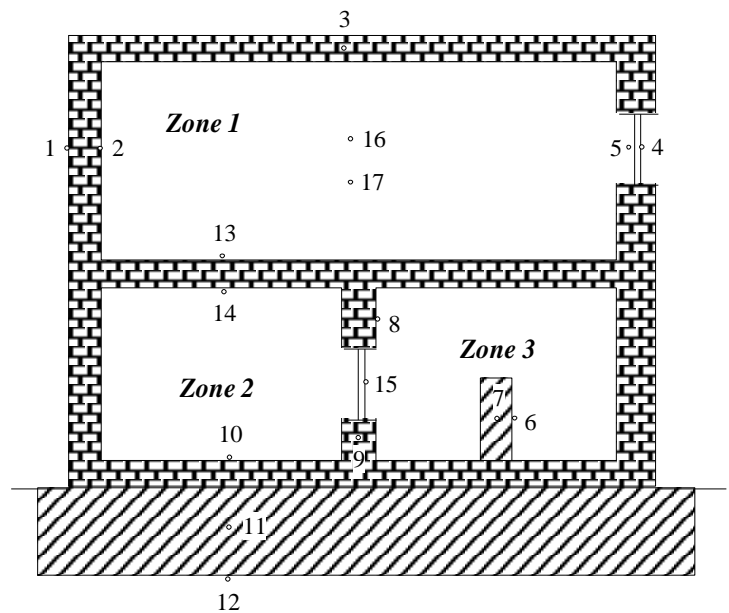

*Fig. 12: Types of nodes.*

```
nodes' structure        =
        {
         ...
         i , ..., 8, ... 1, 2, ..., j,   K1 ,  0   , 0, 1,..., 4, 1 ,...
         j , ..., 9, ... 1, 2, ..., k,   K2 ,  C1  , 0, 0,..., 5, 2 , ...
         k , ..., 9, ... 1, 2, ..., l,   K3 ,  C2  , 0, 0, ...,6, 3 , ...
         l , ..., 8, ... 1, 2, ..., k,   K3 ,  0   , 1, 0, ...,7, 4 , ...
         ...
        }
```

incremental number
    absolute type
        zones' numbers
      following node number
          thermal conductance
            capacitor value
              type relative to zones
                relative zone numbers

*Fig. 13 : Interzone partition and associated data structure*

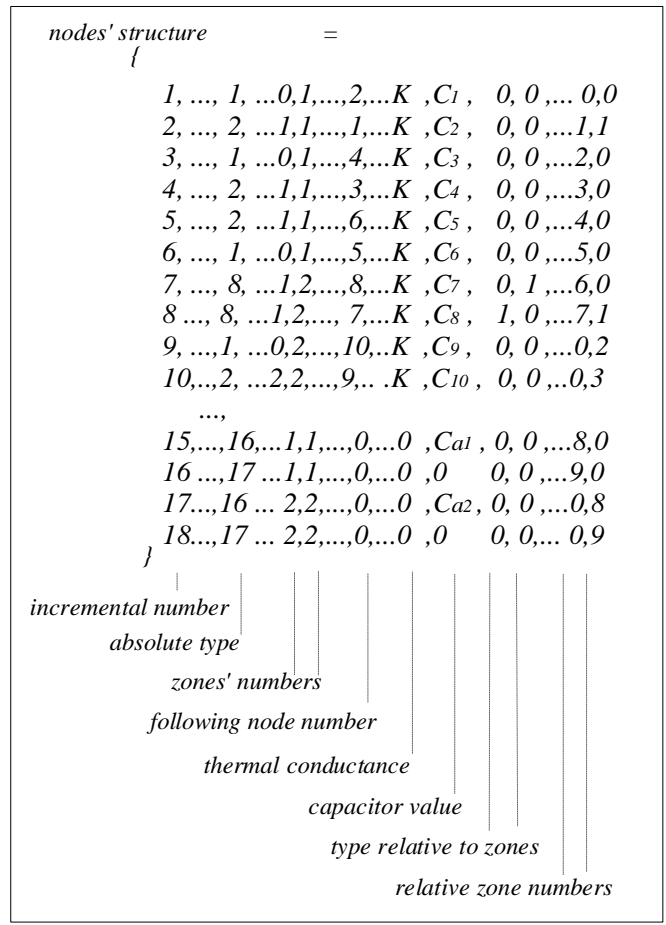

*Fig. 14 : Nodes' structure corresponding to fig. 9 building*